\documentstyle[12pt,psfig,twoside]{article}
\begin{document}
\thispagestyle{empty}
\begin{center}

{\LARGE \bf IX Russian-Finnish Symposium }

{\LARGE \bf   on Radio Astronomy }

\vspace*{5cm}
{\huge \bf ``Multi-Wavelength Investigations

of Solar and Stellar Activity  and Active Galactic Nuclei''}

\vspace{0.5cm} \hspace{0.5cm}
\psfig{figure=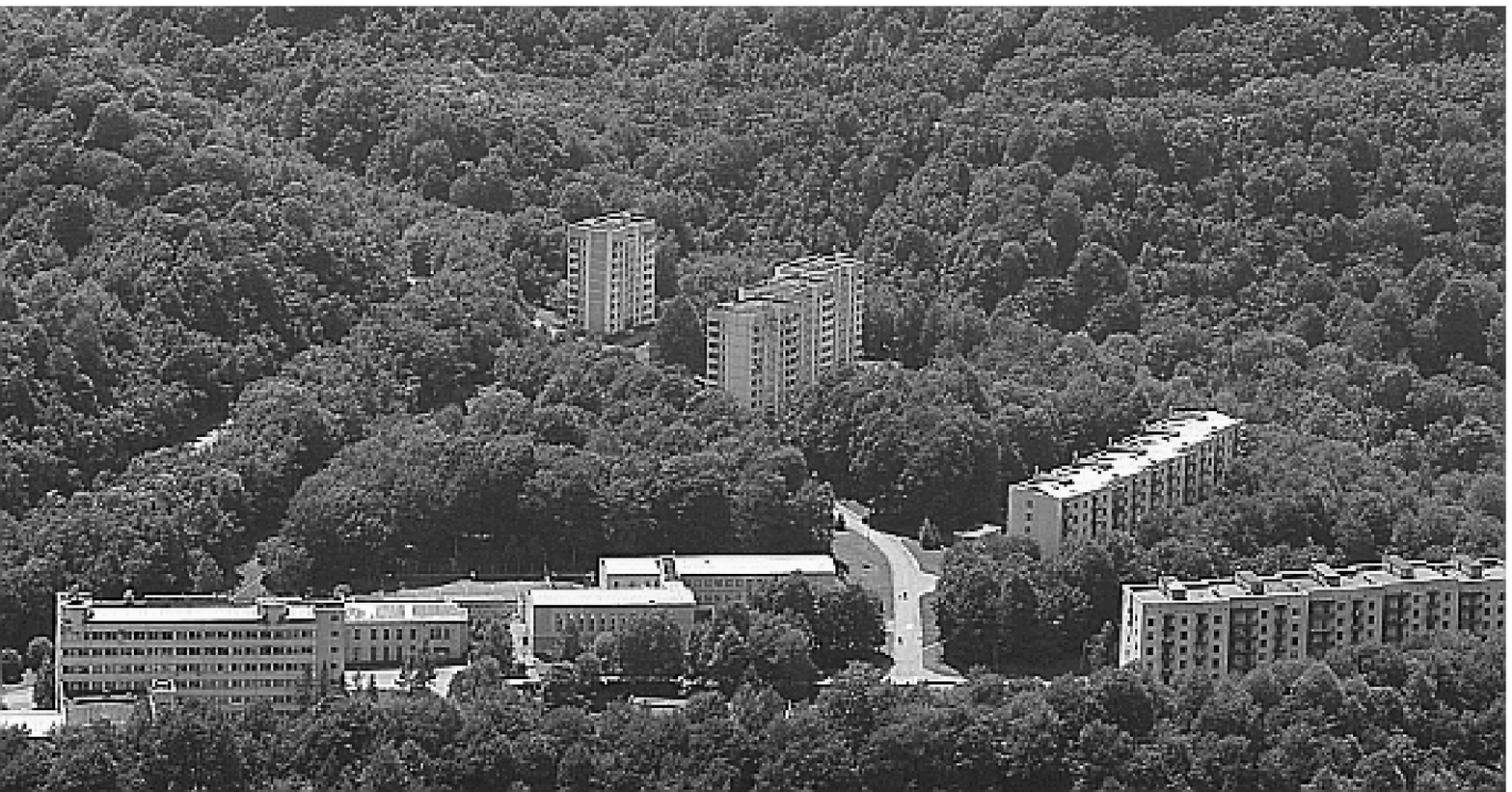,width=12cm,angle=0}

\vspace*{1cm}
{\LARGE \bf Program and abstracts}

\vspace*{5cm}

{\Large \bf 15--20 October 2006, Nizhnij Arkhyz }
\end{center}

\newpage
\thispagestyle{empty}
{
\large
{\bf Scientific Organizing Committee}\\

Marat Mingaliev, chair, Yury Parijskij, \\
Aleksander  Stepanov, Merja Tornikoski, Esko Valtaoja

\vspace*{3cm}

{\bf Local Organizing Committee} \\

Sergei Truskhin (chair), Vladimir Bogod, Ekaterina Filippova, \\
Larisa Martynova,  Julia Sotnikova, Larisa Opeikina, Abdulah Uzdenov

\vspace*{3cm}

{\bf The organizers of the Symposium} \\

Central Astronomical Observatory at Pulkovo \\
Pulkovo chaussee 65/1, Saint-Petersburg, 196140

Special Astrophysical Observatory RAS, \\
Nizhnij Arkhyz, Karachaevo-Cherkassia, 369167

\vspace*{3cm}

{\bf Sponsoring Institutions} \\

Russian Academy of Sciences

Special Astrophysical Observatory of RAS

Central Astronomical Observatory at Pulkovo

Russian Foundation of Basic Research (RFBR)

\newpage
\thispagestyle{empty}
\vspace*{10cm}
\begin{center} {\LARGE \bf ABSTRACTS} \\
(alphabetically ordered)
\end{center}
}
\newpage
\setcounter{page}{9}

\begin{center}
{\large\bf PECULIARITIES OF QPOS OF MICROWAVE EMISSION OF THE FLARING SOLAR ACTIVE REGIONS
}\\[3mm]
{\sl V.E. Abramov-Maximov, G.B. Gelfreikh }\\[2mm]
{  Central Astronomical Observatory at Pulkovo,
Saint-Petersburg, Russia\\
  gbg@saoran.spb.su}\\[2mm]
\end{center}
Quasi-periodic oscillations of solar coronal structures are
effectively registered in particular at microwave waves.
The physical nature of periodical processes in the
corona is clearly connected with  plasma structures of the
solar atmosphere.

On the other hand, the modern conception is that processes leading to
eruption of energy in solar flares are results of energy accumulation
in the corona or chromosphere produced by some reconstruction of
plasma structures.

In spite of developments of the modern
observations predominantly using cosmic techniques, we are still
far from a satisfactory observational understanding of  the
above processes. The QPOs of  solar active regions were found  decades ago
and their connections with the flares were detected. Nevertheless,
the physical  nature and possible forecasting applications are
still far from a reasonable level of knowledge. New development of
microwave instruments with a high spatial resolution and regular
observations (NoRH, RATAN--600, SSRT at Badary) opened a new era
in such studies. Oscillations with periods from few minutes to
hours were investigated and their time variations were specially
analysed. So, we may conclude that investigations of these
parameters for ARs with different levels of flare activity, time
variations including, may lead to a better understanding of the
physics of the problem. Some preliminary results of such studies
are presented in this report, based on an analysis  of  the
Nobeyama radio maps of the Sun with the 10 sec averaging and
covering periods of hours of observations (dates 11 Sep 2001, 07 Oct 2002,
including),  and comparing these with spectral
parameters of the regions obtained with RATAN--600.\\ This study
was partially supported  by the  Program of the Presidium or the
Russian Academy of Sciences.
\vspace{1cm}

\begin{center}
{\large\bf A LINK BETWEEN  VARIABLE OPTICAL CONTINUUM\\ AND RADIO EMISSION OF A COMPACT JET IN THE RADIO-LOUD SEYFERT GALAXY 3C390.3} \\ [3mm]
{\sl  T.G. Arshakian$^1$, A.P. Lobanov$^1$, V.H. Chavushyan$^2$ ,
A.I. Shapovalova$^3$,  J.A. Zensus $^1$,  N.G. Bochkarev$^4$,
A.N. Burenkov$^3$} \\[2mm]
{$^1$Max-Planck-Institut fur Radioastronomie, Auf dem Hugel 69,
53121 Bonn, Germany,\\
$^2$Instituto Nacional de Astrof{\'i}sica, Optica y Electr\'onica,
Apartado Postal 51. C.P. 72000. Puebla, Pue., M\'exico,\\
$^3$Special Astrophysical Observatory of RAS, Nizhnij
Arkhyz, 369167, Russia,\\
$^4$Sternberg Astronomical Institute, University of Moscow,
Universitetskij Prospect 13, Moscow 119899, Russia \\
tigar@mpifr-bonn.mpg.de } \\[2mm]
\end{center}
We present an observational evidence for a relation between
variability of radio emission of the compact jet, nucleus
optical continuum emission and ejections of new jet components in
the radio galaxy 3C390.3. We combine  results from  the monitoring of
3C390.3 in the optical region (Shapovalova et al. 2001; Sergeev at al. 2002)
with ten very long baseline
interferometry (VLBI) observations of its radio emission at
15 GHz carried  out from 1992 to 2002 using the VLBA (Kellermann et al. 2004).
For ten VLBA images, we identified five  moving components (C4-C8) and two
stationary components (D,S1). Proper motions of the moving
components correspond to apparent velocities from 0.8c to 1.5c. No
significant correlation exists for the moving features between
optical continuum and radio emission. However the variations of
optical continuum are correlated with radio emission from a
stationary feature (S1) in the jet. The optical emission follows
radio flares with the mean delays t(S1-opt)$\approx$0.4~year.
Most probably the optical continuum is produced near
the location of radio emission of the S1 stationary component. The
localization of the source of optical continuum with the innermost
part of the jet near S1 implies that the broad line emission
originates in a conical region (dimension $\approx$ 100 light
days)  at a distance of $\ge$ 0.4 pc from the central engine. For
the components C4-C7, the epochs t(S1) of separation from the
stationary feature S1 are coincident, within the errors, with
maxima in optical continuum. This suggests that  radio ejection
events of the jet components are coupled with the long-term
variability of optical continuum.

We suppose that the broad emission lines having a double-peaked
structure originate in two kinematically and physically different
regions of 3C390.3:

1. BLR1 -- the traditional BLR (Accretion Disk (AD) and the
surrounding gas). It is at the distance $\approx$ 30 light days
from the nuclei (Shapovalova et al. 2001).

2. BLR2 -- in a subrelativistic outflow surrounding the jet in the
cone within $\approx$ 100 light days at a distance of $\ge$
0.4 pc from the central engine.

During the nucleus maximal brightness periods most of the
continuum variable radiation is emitted from the jet and ionizes
the surrounding gas, creating a BLR2 that mainly determines the
broad line emission. During the brightness minima the jet
contribution to the ionizing continuum is decreasing and the main
broad line emission comes from the ``classical''\ BLR1 (AD), ionized
by nuclei continuum related with the acctretion at BH. Such a
scenario explains two maxima ($\approx$ 30d and $\approx$100d)
found in the cross-correlation function describing the time-lag of
broad line variations relatively to continuum on the base of the
results of 3C390.3 optical monitoring in 1996-2000 (Shapovalova et al. 2001).

{\it Acknowledgements.} This work was supported by grants: CONACYT
 39560F (Mexico), INTAS (N96-0328) and
RFBR (00-02-16272; 03-02-17123 and 06-02-16843, Russia).
\\[3mm]\indent
{\bf References\\[1mm]}
Shapovalova et al.: {\it A\&A}, 2001, {\bf 376}, 775. \\
Sergeev et al.: {\it ApJ}, 2002, {\bf 576}, 660; \\
Kellermann et al.: {\it ApJ}, 2004, {\bf 609}, 539.
\vspace{1cm}

\newpage
\begin{center}
{\large\bf RATAN-600 MICROWAVE SPECTRAL OBSERVATIONS OF THE SUN -- TODAY AND FUTURE}\\[3mm]
{\sl V.M. Bogod}\\[2mm]
{Special Astrophysical Observatory of RAS, Saint-Petersburg,  Russia\\
  vbog@gao.sb.ru}\\[2mm]
\end{center}
Problems of solar radio emission study with the help of the
RATAN-600 radio telescope are considered. It is shown that the main
scientific results were attained due to the continual improvement
of instrumental parameters. In the case of solar investigations
many new features of active solar plasma were detected, in particular:
\begin{itemize}
  \item A small-scale radio emission structure of the Quiet Sun (a so-called
``radio granulation'');
  \item Neutral line associated sources in solar active regions;
  \item Non-thermal radio emission above sunspots groups, so named ``the decimetre halo'';
  \item Cyclotron lines in the active regions;
 \item  Multiple polarization inversions in flare-productive active regions;
 \item Relations between the double polarization inversions effect at
microwaves and Noise Storms activity at meter waves;
 \item The short-wave increasing of polarization flux before big flares;
 \item The ``darkening'' radio emission effect in flare-productive active
regions;
\item An evolution of polarization flux spectrum before
big flares;
\item Discovery of micro-bursts and their relation with Noise Storms;
\item Detection of a frequency boundary between S- and B-components and others.
\end{itemize}

Examples of the features listed above are described.
A possible future development of solar
investigations with RATAN-600 is discussed.
\vspace{1cm}

\begin{center}
{\large\bf MULTIWAVE RATAN-600 OBSERVATIONS OF POST-ERUPTIVE PROCESSES ON THE SUN}\\[3mm]
{\sl V.N. Borovik$^1$, V.V. Grechnev$^2$, V.E. Abramov-Maksimov$^1$, I.Y. Grigorieva$^1$,
     V.M. Bogod$^3$, V.I. Garaimov$^3$, T.I. Kaltman$^3$, A.N. Korzhavin$^3$}\\[2mm]
{$^1$Central Astronomical Observatory of RAS,
   Saint-Petrsburg, Pulkovskoje shosse, 65, \\
$^2$Institute of Solar Terrestrial Physics, Irkutsk,\\
$^3$ Special Astrophysical Observatory of RAS, Saint-Petrsburg, Pulkovskoe schosse, 65, Russia \\
  borovik@saoran.spb.su}\\[2mm]
\end{center}
During two last decades, authors of some studies of post-eruptive arcades
repeatedly came to the following unexpected conclusions. Microwave emission
of arcades was excessively polarized, presumably due to contribution of
non-thermal electrons. Their lifetime was much longer than the estimated
cooling times, presumably due to the post-eruptive energy release. Finally,
the plasma pressure exceeded the magnetic pressure ($\beta\ge1$) in their hot
top parts. CORONAS-F/SPIRIT observations in the high-temperature ($\sim$10~MK)
line MgXII and multi-wave RATAN-600 observations along with data from other
spectral domains provided important information to
verify these conclusions and assumptions. All above facts were confirmed
in analyses of this data set. They were explained in terms of the standard
flare model (``CSHKP'') elaborated by Yokoyama \& Shibata (1998) to
qualitative account for the chromospheric evaporation, but applied to late
post-eruptive phase. In this case, high $\beta$ conditions indicate magnetic
reconnection
processes responsible for the prolonged heating and particle
acceleration. This approach allows to reconcile the listed facts with known
estimates of parameters of the coronal plasma in post-eruptive arcades, and
to remove seeming contradictions with habitual conceptions.
We consider long-lived post-eruptive arcades observed on
22 October 2001, 2 November 1992, and 28--30 December 2001 and
demonstrate that these
conclusions are valid, because high-density hot regions in their top
parts (thus, high $\beta$ regions) existed for a long time, and their radio
emission contained non-thermal component, which is indicative of the
presence of accelerated particles.
\\[3mm]\indent
{\bf References \\[1mm]}
Yokoyama T., Shibata K.: 1998, {\it ApJ}, {\bf 494}, L113.

\vspace{1cm}

\begin{center}
{\large\bf SPECTRAL MONITORING OF NGC 5548 IN 1996 -- 2004}\\[3mm]
{\sl A.N. Burenkov$^1$, A.I. Shapovalova$^1$, N.G. Bochkarev$^2$,
V.H. Chavushyan$^3$, S. Collin$^4$, V.T. Doroshenko$^2$,
 L. Popovi\'c$^5$, N. Borisov$^1$, L. Carrasco$^3$,
 D. Ili\'c$^6$,  J.R. Valdes$^3$, V.V. Vlasuyk$^1$,
V.E. Zhdanova$^2$}\\[2mm]
{$^1$Special Astrophysical Observatory of RAS, Nizhnij
Arkhyz, 369167, Russia\\
$^2$Sternberg Astronomical Institute, University of Moscow,
Universitetskij Prospect 13, Moscow 119899, Russia\\
$^3$Instituto Nacional de Astrof{\'i}sica, Optica y Electr\'onica,
Apartado Postal 51. C.P. 72000. Puebla, Pue., M\'exico\\
$^4$LUTH, Observatoire de Paris, Section de Meudon, Place Janssen,
92195, Meudon, France\\
$^5$Astronomical Observatory, Volgina 7, 11160 Belgrade, Serbia \\
$^6$Department of Astronomy, Faculty of Mathematics,
    University of Belgrade, Studentski trg 16, 11000 Belgrade, Serbia}\\[2mm]
\end{center}
We present results of a spectral monitoring program of the Seyfert galaxy NGC 5548
with the 6m and 1m telescopes of SAO (Russia) and with the 2.1m telescope
of Guillermo Haro Observatory at Cananea, M\'exico.
 Spectra were obtained with long-slit spectrograph, covering the spectral
range $\sim$(4000--7500)\,\AA \/ with a (4.5--15)\,\AA\,resolution. We found that:

-- Both the flux in the lines and continuum gradually decreased, reaching
minimum values during
May-June 2002. In the minimum state, the wings of H$\beta$ and
H$\alpha$ became extremely weak, corresponding to a Sy1.8 type,
not to a Sy1, as observed previously when the nucleus was brighter.

-- When the line profiles were decomposed into variable and
constant components, the variable broad  component is well
correlated with the continuum variation. It consists of a double
peaked structure with radial velocities of $\sim\pm$1000~km/s
relative to the narrow component. A constant component, whose
presence is independent of the continuum flux variations, shows
only narrow emission lines. The mean, rms, and the averaged over
years, observed and difference line profiles of H$\beta$ and
H$\alpha$ reveal the same double peaked structure at the same velocities.
 The relative
intensity of these peaks changes with time. During 1996, the red
peak was the brightest, while in 1998--2002, the blue peak
became the brighter one. Their radial velocities vary in the range
$\sim$500--1200 ~km/s.

-- In 2000--2002 a distinct third
peak appeared in the red wing of H$\alpha$ and H$\beta$ line
profiles. The radial velocity of this feature decreased between
2000 and 2002: by the observed profiles, from $\sim$ +(2500--2600)\, km/s
to $\sim+2000$\, km/s and it is clearly seen on the
difference profiles.

-- The fluxes of various parts of the line
profiles are well correlated with each other and also with the
continuum flux. The blue and red parts of the line profiles at
the same radial velocities vary in an almost identical manner.

-- Our results favor the formation of the broad Balmer lines
 in a turbulent accretion disc with large and moving
``optically thick'' inhomogeneities, capable of reprocessing
the central source continuum.

We made an attempt to investigate the variability of physical parameters in
the 
{\bf BLR} of
NGC 5548 using the Boltzmann plot method given by Popovi\'c (2003).
We applied the  method on the broad Balmer lines, and found that
variability seen in lines is also present in the electron
temperature ({\bf T}). We found that the average 
{\bf T} for the  considered period was
$\approx$ 10000K, and that it varies from 6000K (in 2002)
till  15000K (in 1998). This
variation correlates with the optical continuum flux ($r=0.85$) and
may indicate existence of an accretion disk in the {\bf BLR} of NGC 5548.
We found that Partial Local Thermodynamical Equilibrium approximation
is valid for at least one part of the BLR of NGC 5548.

The detailed discussion of these results is done in our papers (Shapovalova at al. 2004;
Popovi\'c et al. 2005).

{\it Acknowledgements.}
This work was supported by grants from CONACYT 39560-F (Mexico),
 RFBR 06-02-16843 (Russia) and the Ministry of Science and Environment
Protection of Serbia.
\\

{\bf References\\[1mm]}
Popovi\'c L. \v C.: {\it ApJ}, 2003, {\bf 599}, 140. \\
Popovi\'c L. \v C. et al.: {\it Astro-ph/0511676}, 2005 (in press). \\
Shapovalova A. I. et al.: 2004, {\it A\&A}, {\bf 422}, 925.
\vspace{1cm}
\newpage
\begin{center}
{\large\bf MEAN SPECTRAL INDEX OF THE FAINTEST NVSS OBJECTS FROM RZF DATA }\\[3mm]
{\sl N.N. Bursov$^1$, M.L. Khabibullina$^2$}\\[2mm]
{$^1$Special Astrophysical Observatory  of RAS, Nizhnij Arkhyz, 369167, Russia\\
$^2$Kazan State University, Kazan, Russia \\
  nnb@sao.ru}\\[2mm]
\end{center}
A spectral index of a 'mean' radio source from NVSS catalog
was estimated by summarizing of the drift scans of the RZF survey.
In the area of RZF servey (RA2000:0-24; DEC2000:40.5-42.5) there are 28000
NVSS sources with a total flux density from 2 to 400 mJy at 1.4 GHz.
These NVSS sources are considered as point ones.
We have divided this flux range to 12 bins: 2--4, 4--6, 6--8, 8--10, 10--13,
13--17, 17--24, 24--35, 35--60, 60--110, 110--220, 220--400 mJy.
We summarized 7500 intercepts of one-hour records at 4 GHz of RZF survey by
a number of NVSS sources in each bin.
Indeed we 'detected' a 'avarage' source in each bin of NVSS sample and
obtained 'average' flux densities at 4 GHz for each flux bin:
0.9, 1.7, 2.6, 3.7, 4.3, 6.9, 7.6, 10.5, 17.2, 27.5, 60.0,
177 mJy respectively.

The data give a rough estimate of the 'average' spectral indices (1.4--4~GHz)
for each NVSS sources bin.  We found that an 'average' spectral
index is a function of NVSS flux densities.
A comparison with expected mean spectral indices and the reliability of
such a estimate are discussed.

However, in the bin of 3--8 mJy the spectra have spectral index equal
to --0.8, i.e. the relative fraction of the steep spectrum sources
decreased probably because a number of FRII radio sources decreases just in
this flux range (Jarvis \& Rawlings, 2004).
\\[3mm]\indent
{\bf References\\[1mm]}
Jarvis M.J., Rawlings  S.:
   {\it New Astron.Rev.}, 2004, {\bf 48}, 1173.
\vspace{0.7cm}

\begin{center}
{\large\bf RZF SURVEY}\\[3mm]
{\sl N.N. Bursov, E.K. Majorova, T.A. Semenova,
M.G. Mingaliev, A.B. Berlin, N.A.~Nizhel'skij, P.G. Tsibulev }\\[2mm]
{Special Astrophysical Observatory of RAS, Nizhnij Arkhyz, 369167, Russia\\
  nnb@sao.ru}\\[2mm]
\end{center}
The completed second version of the RZF catalog at 7.6 cm wavelength is presented.
A reality of faint and new objects of the catalog is evaluated.
Radio spectra of NVSS objects are plotted and a new distribution of
spectral indices is found.

New information about a contribution of background of the faint radio sources
with inverted spectrum and about a ``sub-mJy'' population of the background  radio
sources in the cm range is found.
It is shown that an average spectral index of radio sources at the mJy level is
flatter, a percentage of classical  radio  galaxies  FRII type drops,
but the population of objects with inversion spectra is negligible.

Candidates of the most distant radio galaxies ($z > 4-5$)
catalog are selected by colors from the SDSS counterparts.

A decrease  of white noise of radiometers is achieved
by a large integration time at pixel that  is below the level of
WMAP experiment.

The synchrotron component of the foreground  Galaxy emission with
high accuracy at earlier unstudied angular scales was removed.
\vspace{1cm}

\begin{center}
{\large\bf CONTINUUM AND BROAD EMISSION LINE VARIABILITY OF SEYFERT GALAXIES}\\[3mm]
{\sl V.T.\,Doroshenko$^{1,2}$, S.G.\,Sergeev$^2$,
V.I.\,Pronik$^2$}\\[2mm] {$^1$\it{Crimean Lab. of the
Sternberg Astronomical Institute, Moscow University, Russia\\}}
{$^2$\it{Scientific-Research Institute, Crimean
Astrophysical Observatory, Ukraine;}}\\
  p/o Nauchny, 98409, Crimea, Ukraine\\
  {vdorosh@sai.crimea.ua}\\[2mm]
\end{center}
We focused on the observed properties of some
Seyfert galaxies without detailed discussion of underlying
physical mechanisms. Our purpose was to show a diversity of
observed effects due to variability in spectra of six galaxies
(NGC~4151, NGC~5548, Mrk~6, Ark~120, 3C~390.3, Arp~120B), optical
spectral and photometric monitoring of which was carried out in
the Crimean Astrophysical Observatory and Crimean Laboratory of
the Sternberg Astronomical Institute over many years. This
monitoring shows that
\begin{itemize}
\item All light curves demonstrate a variability on different
time scales from days to years.
\item Amplitude of variations increases with increasing of
the time interval of observations.
\item The flux in emission lines changes in response to the flux
variation of the ionizing continuum source with some time delay.
Thus, the emission lines ``echo''\ or ``reverberate''\  the continuum
changes. This time delay is due to  light-travel time effects
within the BLR.
\item The time delay of the broad H$\beta$ emission line flux relative
to optical continuum in the vicinity of the H$\beta$ line lies
in an interval from 9.2 days for Arp~102B to 80 days for 3C~390.3.
This means that a region of the most effective emission in the
H$\beta$ and H$\alpha$ lines is fairly small, and it is located at
a distance of about 9 -- 80 light days from the continuum source.
\item We found one very strange and inexplicable case
(3C~390.3) when a lag calculated from the broad H$\alpha$ line
significantly exceeds that of the H$\beta$. In all other cases,
the H$\alpha$ and H$\beta$ lines have a similar lag.
\item We revealed a slightly different lag for two time
intervals for NGC~5548: $\sim$26 days in 1972--1988 and $\sim18$
days in 1989--2001.
\item Analysis of the lag as a function of the radial velocity does
not show any reliable evidence of the pure radial dominated outflow,
possibly, except for NGC~4151.  A weak evidence of radial inflow was
found in Mrk~6 and Ark~120. Predominantly, we found that there is
no pure radial inflow or outflow in the BLR of the considered
galaxies. In all cases, the kinematics mainly looks like a chaotic
or rotational motion.
\item We also found that the lag for the central part of the broad
H$\beta$ emission line is slightly larger than for wings (NGC~5548).
\item The lag slightly increases with increasing of the continuum
flux (Ark~120). This fact is consistent with a virial relation
between the velocity field and the distance of the emitting
region: the velocity field diminishes with increasing distance
from the central continuum source. This implies that the velocity
field is dominated by a central massive object.
\item Not only the flux of broad emission line but also the line profiles
appreciably changed with time. The emission-line profile changes
usually occur on a time scale that is much longer than the
light-travel time scale.
\item In all cases, the excess between the normalized Balmer
profiles and the mean normalized profile shows a very complicated
behavior both over time and wavelength, and it can hardly be
related to the expected reverberation signal from the simple disk
model. The profile evolution for some galaxies (NGC~4151, Mrk~6,
3C~390.3) can be reproduced to larger or smaller extent with the
two-component model in which profile changes are due to changes in
the relative strength of two variable components with a fixed
shape. The double peaked profile was often observed among the
discussed objects. Profile decomposition gives one component that
dominates in the central part of the profile, while the double
peaked component dominates in the profile wings. However, the
moving features of the profile shapes observed, e.g., in 3C~390.3,
NGC~4151, and Arp~102B can be a result of rotating redistribution
of matter in the Keplerian disk.
\end{itemize}

{\it Acknowledgements.}The research  was made in partly by the award
UP 1-2549-CR-03 of the US Civilian Research and Development
Foundation for the Independent States of the Former Soviet Union
(CRDF) and by the Russian Foundation for Basic Research (RFBR) grant
06-02-16843.
\\[3mm]\indent
\vspace{1cm}

\begin{center}
{\large\bf THE STUDY OF A MAGNETIC FIELD STRUCTURE IN BLAZAR JETS BY OPTICAL PHOTOPOLARIMETRIC DATA}\\[3mm]
{\sl Yu.S. Efimov$^1$, E.Yu. Vovk$^2$ }\\[2mm]
{$^1$RSI ``Crimean Astrophysical Observatory'', Nauchny, Crimea, 98409, Ukraine\\
 $^2$National State University, Kiev, 01033, Ukraine\\
  Yuri.Efimov@mail.ru}\\[2mm]
\end{center}
There are no direct methods to measure magnetic field intensity in
sources of synchrotron radiation. Usually estimations of magnetic
field intensity are obtained from the Faraday rotation measure in radio region
with additional data about a path length and an electron density. Besides the
Faraday rotation of polarization plane there are observational parameters
which are connected directly with the intensity of magnetic field
and its structure: a polarization degree and a spectral index. Both these
parameters can be obtained from optical observations. On the other
hand, various theoretical models connect magnetic field intensity
with polarization degree.\\  From two polarimetric modes observed in
blazars: circular and linear, the last one is rather large (from 1\%
to 40\%) and can be detected easily in contrast to the circular
polarization which is very small (fraction of percentage). In its turn,
the linear polarization is directly connected with spectral index. So,
the comparison of theoretical predictions from various models of magnetic
field with observed dependencies between polarization degrees and spectral
indices allows to significantly confine an uncertainty of the magnetic
field intensity estimations. The first such attempt had been made about 30
years ago by Nordsieck (1976). Moreover, such a comparison allows to follow
the variations of the magnetic field intensity from observed variations
of polarization degree and spectral index.\par In this report we present
a result of such comparison for two well-known blazars 3C~66A and OJ~287.
It was shown that in both sources the decrease of polarization
and spectral index indicate: an increasing disorder of magnetic fields
of relativistic jets.
\\[3mm]\indent
{\bf References\\[1mm]}
Nordsieck K.H.: {\it ApJ}, 1976, {\bf 209}, 653. \\
\vspace{1cm}

\begin{center} 
{\large\bf THE SUPERCRITICAL ACCRETION DISK IN SS433 AND ULTRALUMINOUS X-RAY SOURCES   }\\[3mm]
{\sl S. N. Fabrika }\\[2mm]
{Special Astrophysical Observatory of RAS, Nizhnij Arkhyz, 369167, Russia \\
  fabrika@sao.ru}\\[2mm]
\end{center} 
SS433 is the only known persistent supercritical accretor, it may
be very important for understanding ultraluminous X-ray sources (ULXs)
located in external galaxies. We describe  main properties of the SS433
supercritical accretion disk, jets and its radio nebula W50. \\
Basing on observational data of SS433 and published 2D simulations of
supercritical accretion disks we estimate parameters of a funnel in
the disk/wind of SS433 and discuss formation of jets and a nebula.
Critical observations which may throw light upon  nature of ULXs come from
nebulae observations around ULXs.
We present results of 3D-spectroscopy of nebulae around
several ULXs located in galaxies at distances of 3-6 Mpc. We found that
nebulae are powered by their central black holes. The nebulae are
shocked and dynamically perturbed probably by jets.
\vspace{1cm}
\newpage

\begin{center} 
{\large\bf RESULTS OF THE PROMINENCE OBSERVATIONS AT MICROWAVES DURING THE MAXIMAL PHASE OF THE TOTAL SOLAR ECLIPSE OF MARCH 2006 }\\[3mm]
{\sl O.A. Golubchina, V.M. Bogod,  A.N. Korzhavin, N.N. Bursov, S. Kh. Tokhchukova  }\\[2mm]
{  SPb branch of SAO RAS, St.Petersburg,  Russia\\
  oag@OG4466.spb.edu}\\[2mm] 
\end{center} 
Results of the prominence radio emission study according to observations of
the solar eclipse on 2006 March 29  with the Northern sector
and with the Southern sector with the Flat mirror of RATAN-600 are
discussed. Investigation of the prominence located in the NE
solar limb is executed in the (1.03-5.0) cm wavelength range.
These observations are unique because
the solar eclipse was observed with one telescope simultaneously by two
different methods. This enables to supplement mutually the data 
received by observations with two different sectors and gives
an opportunity to control some obtained results. The angular
resolution of the antenna in the horizontal direction is from 
17.5~arcsec up to 47~arcsec in the (1.88-5.0) cm wavelength range 
with the Southern sector and the Flat mirror and from 0.44~arcmin to
1.68~arcmin in the (1.03-3.9)~cm range with the Northern sector of
the RATAN-600. An average angular size of the prominence source
in the specified wavelength range is about 30~arcsec.
From observations with the Northern and Southern sectors, the position of
maximum of the prominence radio source have been found to coincide
with the prominence top of the solar image in the He~II~304~line
(SOHO,~$\phi$=$45^{o}$, NE limb of the Sun). The radio fluxes of 
the prominence were obtained in the wavelength range from 1.03
to 5.0 cm. The fluxes in the range $\lambda$~=~1.38~-~4.0(cm) are 
equal to F($\lambda$)=0.8-0.01(s.f.u.). These values coincide for observations
with both sectors of RATAN-600, meanwhile the methods of the observations and
the techniques of data processing of the observations with two 
sectors were different.
 
The obtained spectrum of the prominence
defined a thermal mechanism of the prominence
radio emission in the (1.03-5.0) cm range. There is a sharp 
decrease of the prominence radio flux down to value F~=~0.02~s.f.u.
in comparison with expected value for the 1.03~cm wavelength, 
according to received dependence F~($\lambda$). Possibly it is 
caused by the Moon closing the prominence at the moment of 
observation. A relative position of the Moon and the Sun at the 
moment of the solar eclipse maximum phase allowed to estimate
the height of the prominence above the photosphere. The derived 
brightness temperatures of the prominence are equal to 
$T_b~=~(5450-17900)^{o}$~K in the (1.84-4.21)~cm wavelength 
range. It was registered a bipolar structure of the radio source 
associated with the prominence. The degree of circular 
polarization of the source is: $p=~(5-10)\%$ for the 
(1.84-5.0)~cm wavelengths. \\
{\it Acknowledgements.} This work is supported by the RFBR grant N05-02-16228.
\vspace{1cm}
\newpage
\begin{center}
{\large\bf  EXTRAGALACTIC SOURCE VARIABILITY  STUDIES OF COMPLETE SAMPLES WITH RATAN-600}\\[3mm]
{\sl A.G. Gorshkov$^1$, M.G. Mingaliev$^2$}\\[2mm]
{$^1$Sternberg Astronomical Institute, Moscow State
University, Universitetskij pr., 13, Moscow, 119992, Russia}\\
{ $^2$Special Astrophysical Observatory of
RAS, Nizhnij Arkhyz,  369167, Russia}\\
algor@sai.msu.ru\\[2mm]
\end{center}

We present preliminary results of the study and analysis of complete
and limited in flux density source samples from the Zelenchuk and MGB surveys
carried out with the RATAN-600 radio telescope.

This work pursues two the main aims:

\begin{itemize}

\item{ Investigation of general statistical characteristics of
a discrete source sample;}

\item{ Investigation of  variability processes in discrete sources in
a wide range of wavelengths. That assumes a study of the
amplitude and frequency characteristics of variability on all time
scales from several years to several days.}

\end{itemize}

The following results of this investigation have been revealed (or
confirmed):

\begin{enumerate}

\item{The long-term variability with a time-scale more than several
years. Our observations have allowed to trace
a complete evolution cycle of the isolated flare in a number of sources
from its occurrence before decay and to specify its amplitude
and frequency characteristics. There is no plateau at the
maximum of a flare, a flare increase and decay are well
described by a exponential temporal law.
Such a form of  flare is not described by any model.}

\item{The variability with a time-scale about tens of days.
We consider the variability with a week time scale as an important
result of our investigation. It is a new type of the variability and, as it was
found out, it is widespread enough. Approximately 10\% of sources with
flat spectra have a similar variability. The variability often has
a cyclic nature and, basically, is exposed at frequencies below 10 GHz.}

\item{The existence of variability with a time-scale about four days
for all compact radio sources. We are sure that the given type of variability is not
a property of a radio source, but a result of radiation propagation in  inhomogeneous interstellar medium.}
\end{enumerate}
\vspace{1cm}
\newpage
\begin{center}
{\large\bf ON THE SKY SPECTROSCOPY WITH RATAN-600}\\[3mm]
{\sl I.V. Gosachinskij}\\[2mm]
{  Saint-Petersburg Branch of the Special Astrophysical observatory of RAS, \\
  196140, Saint-Petrsburg, Pulkovskoje schosse, 65,  Russia\\
  gos@sao.ru}\\[2mm]
\end{center}
Observations of radio lines with the RATAN-600 radio telescope are being
carried out during 30 years in the domain of

-- systematical investigations of a cloudy structure of the Galactic
interstellar medium in the lines of HI at 21 cm, OH at 18 cm, H$_2$CO
at 6.2 cm and H$_2$O at 1.35 cm with the aim of understanding its
structure, dynamics, evolution and interaction with other
Galactic populations;

-- investigation of physical and evolutionary status of large structures
of interstellar gas;

-- a search of gas clouds at cosmological distances with the help of
their probable line emission.

The most interesting results were obtained in the investigation of
HI gas around SNR's and HII regions, recombination line H110$\alpha$
in the Orion nebula, HI Super Shells, a formaldehyde cloud in the source
Sgr B2, statistical characteristics,  ``scale relations''\  and
internal motions of the HI clouds and properties of interstellar gas
at high Galactic latitudes.
\vspace{1cm}

\begin{center}
{\large\bf INVESTIGATION OF INTERACTION BETWEEN SUPERNOVA REMNANTS AND  INTERSTELLAR MEDIUM}\\[3mm]
{\sl I.V. Gosachinskij, A.P. Venger, Z.A. Alferova}\\[2mm]
{  Saint-Petersburg Branch of the Special Astrophysical Observatory of RAS \\
   196140, Saint-Petersburg, Pulkovskoje shosse, 65,  Russia\\
    gos@sao.ru}\\[2mm]
\end{center}
During 1999-2006 the second stage of investigation of HI distribution
around supernova remnants (SNRs) was carried out with the RATAN-600 radio
telescope.
In contrast to a previous stage  during 1985-87 now SNR of large angular
dimension (greater than $10'$) and specific type -- S (shell) were
selected, independently from their radio brightness. It is obvious that
namely such objects have sufficiently large ages for demonstrating a evidence
of interaction between their shock waves and surrounding neutral gas.
Such a evidence could be detected in the ``Right Ascension -- Velocity'' ($\alpha-V$) maps
as ring like structures, whose parameters may bring information about
sizes, ages and energy of supernova explosion.

Now we have observed about 130 SNRs and 105 $\alpha-V$ maps are plotted.
Some interesting objects (such as S147, Cygnus Loop,
HB3) were studied in detail and results were published separately.
Soon all $\alpha-V$ maps will be available at our web-site.
\vspace{1cm}
\newpage

\begin{center} 
{\large\bf LONG-TERM RADIO TIME SCALES OF ACTIVE GALACTIC NUCLEI}\\[3mm]
{\sl T. Hovatta $^1$, M. Tornikoski $^1$, M. Lainela $^2$, E. Valtaoja $^2$, I. Torniainen $^1$, \\M.F. Aller $^3$, H.D. Aller $^3$}\\[2mm] 
{ $^1$ Mets\"ahovi radio Observatory\\ 
Mets\"ahovintie 114 02540 Kylm\"al\"a, Finland\\ 
$^2$ Tuorla Observatory, University of Turku, Finland\\ 
$^3$ Department of Astronomy, University of Michigan\\ 
tho@kurp.hut.fi}\\[2mm] 
\end{center} 
We have studied long-term variability time scales of a large sample
of Active Galactic Nuclei at several frequencies between 4.8 and 230 GHz.
The sample consists of 80 sources from different classes of AGN. In our
sample we have quasars, BL Lacertae objects and Radio Galaxies. Our
sample consists of sources from the Mets\"ahovi monitoring programme
where a sample of compact extragalactic radio sources has been monitored
for over 25 years.  In addition we use lower frequency data from the
University of Michigan monitoring programme and data obtained from the
SEST-telescope between 1986 and 2003.

We used the first order structure function, the discrete auto-correlation
function and the Lomb-Scargle periodogram to study the characteristic time
scales of variability. We were interested in finding differences and
similarities between classes and frequencies. Also the methods were
compared in order to find the most efficient one for different purposes.

We have also compared the results of this study with earlier structure
function analysis by Lainela \& Valtaoja (1993). In the earlier analysis
the structure function was used to study 42 sources from the Mets\"ahovi
monitoring sample at 22 and 37 GHz frequencies. We wanted to find out
how the time scales have changed after the amount of monitoring data
has more than tripled.

The main conclusion of our study is that in these sources smaller
variations happen also in short time scales but larger outbursts only in
time scales of many years. Therefore in order to study how often sources
are in active state and how long these flares typically last the long-term
monitoring is needed.

{\bf References\\[1mm]}
Lainela M., Valtaoja E.: {\it ApJ}, 1993, {\bf 416}, 485.
\vspace{1cm}

\begin{center}
{\large\bf RADIO VARIABLE SOURCES WITH THE RT32 RADIO TELESCOPE}\\[3mm]
{\sl M. Harinov$^1$, S.A. Trushkin$^2$,  A. Mikhailov$^1$}\\[2mm]
{Institute of Applied Astronomy of RAS, SAO RAS, Saint-Petersburg, \\
 Special astrophysical observatory of RAS, Nizhnij Arkhyz,  Russia\\
  kharma78@rambler.ru}\\[2mm]
\end{center}
We discussed first results of radio observations of AGNs and
microquasars with the RT32 radio telescope (Zelenchuk) during 2004-2006.
We carried out  more than 20 sets of observations of the microquasars:
SS433, Cyg X-3 and LSI+61d303 at frequencies 2.3 and 8.45 GHz.
Usually during 1-3 days these sources were observed in a multi-scanning
mode, when the antenna elevation or the antenna azimuth were changed following a
cosmic source.  Thus, for 3-5 daily observations with a duration of 30-60
minutes we integrated up to 100 single scans, which could be used to study a
fast intra-day variability. The flux sensitivity of about 10-20 mJy was
reached at both frequencies.
From November 2004 to August 2006 in
twelve two-day sets of observations the sample of 50 bright variable
extragalactic sources from 3ó and WMAP catalogs and from the sources
list selected for the Russian-Finnish program of AGNs was studied.
It is important for such programs that from March 2005 the RT32 observations
were regularly carried out simultaneously at both frequencies 2.3 and 8.5 GHz
in two circular polarizations. A good agreement of the RATAN and RT32
flux measurements of microquasars and AGNs was obtained.

{\it Acknowledgements.} The authors are thankful to the
RFBR and Presidium of RAS for support by grants. S.T. is very thankful
to  the IAA Program Commitee for regular allocation of the RT32 observation time.
\\[3mm]\indent
{\bf References\\[1mm]}
Trushkin S.A., Harinov M.A., Michailov A.G.: 2005, ATel, {\bf N488}, 1 \\
Trushkin S.A., Pooley G., Harinov M.A., Mikhailov A.G.: 2006, ATel, {\bf 828}, 1 \\
Trushkin S.A., Bursov N.N., Valtaoja E., Nizhelskij N.A., Tornikoski M.,
Mikhailov A.G. Harinov M.:
HEA-2006, Abstract book, Moscow, Dec 24-28 2006.
\vspace{1cm}


\begin{center}
{\large\bf ESTIMATIONS OF EFFECTIVE HEIGHT, SIZE AND BRIGHTNESS TEMPERATURE OF SOLAR CYCLOTRON SOURCES}\\[3mm]
{\sl  A.N. Korzhavin,  T.I. Kaltman}\\[2mm]
{Special Astrophysical Observatory of RAS, Saint-Petersburg, Russia\\
arles@mail.ru}\\[2mm]
\end{center}
The modeling of microwave emission from a spot-associated cyclotron source was
done to  refine the method  of  estimations of effective brightness
temperature, size and   height above photosphere in the processing of RATAN-600
observations.  The simple model of a source  with a dipole distribution
of magnetic field and with a two-step transition region between the cold
dense chromosphere and the  hot corona was used.

When a source approaches the limb the decrease of a source visible
size in E-W direction takes place due to the projection effect, which causes
the decrease of its effective size in processing of a one-dimensional scan
of RATAN-600.  The subsequent Gauss-analysis would overestimate values
of brightness temperatures if necessary  corrections were not done.

The same projection effect leads to the fact, that the size of source
observed in  polarized  emission (Stokes parameter V) exceeds  the size
of the source in full intensity emission (Stokes parameter I)   due
to non circular distribution of polarized emission.

The method of estimates of the effective height of emission above
photosphere level by measurements  of an emission centre of weight
declination   from a source geometrical  centre   at approaching to the
limb was modeled and presented.

This work was supported by the RFBR grants 05-02-16228 and 06-02-17034a.
\vspace{1cm}

\begin{center}
{\large\bf RATAN-600 OBSERVATIONS OF MICROWAVE STRUCTURE  OF THE QUIET SUN}\\[3mm]
{\sl T.I. Kaltman, V.M. Bogod, A.N. Korzhavin, S.Kh. Tokhchukova}\\[2mm]
{Special astrophysical observatory of RAS, Saint-Petersburg, Russia\\
arles@mail.ru}\\[2mm]
\end{center}

To investigate microwave emission of the quiet Sun  the observations
with RATAN-600 from  September, 2005 to March, 2006 in the range  6-16.4
GHz with the 1\% frequency  resolution were used.

We present an analysis of observational data for several days with
different positional angles.  A small-scaled structure with the size of
20-40 arc sec is regularly observed with RATAN-600 one-dimensional observations.
A high degree of correlation for separate elements
of the structure in the different frequencies channels at all band of
the observations exists. Our estimates of  an average life time are
several hours. There is a direct dependence between the sizes and life
of time for separate elements.

The spectra of brightness temperatures grow with wavelength. The
emission polarization is very likely negligible.  The characteristics
of  presented observed structure  are very close to ones of a super
granulation (chromosphere network) which is not sufficiently investigated
in  microwaves. The separated bright sources are identified with bright X-ray
points or bipolar magnetic structure.

Our modeling  demonstrates that the structure of the chromosphere network
can  exist in a wide spatial range, but really only the sources with
the sizes of 20-40 sec of arc can be detected at microwaves. Possible
mechanisms of such radio emission are discussed.

The daily monitoring  with RATAN-600 observations provides
possibilities to regularly estimate a state of the quiet Sun  by  emission
characteristics of  microwave small-scaled structure  and to trace rises
of new  centers of activities.

This work was supported by the RFBR grants 05-02-16228 and 06-02-17034a.
\vspace{1cm}

\begin{center} 
{\large\bf ABOUT A DOUBLE INVERSION SIGN OF POLARIZATION MICROWAVE EMISSION FROM FLARE-PRODUCTIVE ACTIVE REGION}\\[3mm]
{\sl V. Kotelnikov$^1$, V. Bogod$^1$, L. Yasnov$^2$}\\[2mm]
{$^1$Saint-Petersburg branch of Special Astrophysical Observatory of RAS\\
  Pulkovskoe Shosse, 65, Saint-Petersburg,  Russia\\ 
 $^2$Saint-Petersburg State University  \\
198904 Saint-Petersburg, St. Peterhof, Ul'anovskaya street 1 \\
  vasian.spbu@mail.ru}\\[2mm] 
\end{center} 
Polarization inversions have been detected in some microwave 
sources by Piddington et al. (1951) and by Peterova at al. (1974).
Tokhchukova et al. (2002) have shown that a more complex phenomenon
is observed in the flare-active regions:  before a powerful flare
the sign of circular polarization changed twice within a
narrow frequency range. Here we discuss observations of flare-productive
active regions. These observations were carried out with the RATAN-600
radio telescope in a broad radio range for a period from 2000 to 2004.
The double inversion has been observed in several events
before powerful proton flares. We propose two alternative models for
explanation  of this phenomenon. The first model is the existence of the magnetic
``hole'' in active the region and the second model is the propagation of
radio waves through a layer with zero magnetic field.
\\[3mm]\indent 
{\bf References\\[1mm]} 
Peterova N.G., Akhmedov S.B.: {\it Soviet Astronomy}, 1974, {\bf 17}, 168.\\
Piddington J. H., Minnnett H. C.: {\it Austral. J. Sci. Res.}, 
1951 {\bf A4}, 131.\\ 
Tokhchukova S. Bogod V.: {\it Solar Physics.}, 2002, {\bf 212}, 99.
\vspace{1cm}

\begin{center}
{\large\bf RADIO PULSATIONS FROM THE AD LEO FLARE AND ELECTRIC CURRENT DIAGNOSTICS}\\[3mm]
{\sl E.G. Kouprianova$^1$, A.V. Stepanov$^1$, V.V.Zaitsev$^2$\\[2mm]
{$^1$\it{Central Astronomical Observatory at Pulkovo, \\
  Pulkovo Chaussee 65/1 Saint Petersburg 196140, Russia}}\\
{$^2$\it {Instutute of Applied Physics, Nizhny Novgorod, Russia}}\\
  lioka@gao.spb.ru}\\[2mm]
\end{center}
Using pulsations characteristics of AD Leo radio flares observed
by Bastian~et~al.~(1990) with the Arecibo 300m and by Stepanov~et~al.~(2001)
with the Effelsberg 100m radio telescopes the values of electric
currents $(7$--$40)\times10^{11}$~A and plasma parameters in
stellar flares are determined. It was shown that radio pulsations
can be due to both ``sausage'' oscillations as well as current
RLC-oscillations in a flare loop (Zaitsev et~al. 1988, 2004).
Explanation of very intense radio bursts ($T_b \approx 10^{15}$~K)
in terms of coherent plasma emission gives the magnetic field
value (100--300 G) and the electron number density
($10^{10}$--$10^{11}$~cm$^{-3}$) in the flares. The energy of
electric current stored in the flares was estimated as
$(1$--$50)\times 10^{25}$~J. It is shown that $< \; \sim \! 10$\%
of stored energy was released in the flares.
\\[3mm]\indent
{\bf References\\[1mm]}
Bastian~T., Bookbinder~J., Dulk~G.A., Davis~M.: {\it
Astrophys.~J.}, 1990,
{\bf 353}, 265. \\
Stepanov~A.V., Kliem~B., Zaitsev~V.V.~et~al.: {\it
Astron.~Astrophys.}, 2001, {\bf 374}, 1072. \\
Zaitsev~V.V., Stepanov~A.V., Urpo~S., Pohjolainen~S.: {\it
Astron.~Astrophys.}, 1998, {\bf 337}, 887. \\
Zaitsev~V.V., Kislyakov~A.G., Stepanov~A.V., Kliem~B., Fuerst~E.:
{\it Astron.~Lett.}, 2004, {\bf 30}, 319.
\\[3mm]\indent
\vspace{1cm}
\newpage
\begin{center}
{\large\bf NATURE OF ACTIVE GALACTIC NUCLEI FROM MASSIVE INSTANTANEOUS RADIO SPECTRA STUDY WITH RATAN--600 IN 1997--2006 SUPPLEMENTED BY VLBA EXPERIMENTS}\\[3mm]
{\sl
Yu.A.~Kovalev$^1$, Y.Y.~Kovalev$^{1,2}$, G.V.~Zhekanis$^3$,
N.A.~Nizhelsky$^3$
}\\[2mm]
{ $^1$Astro Space Center of Lebedev Physical Institute,\\
  Profsoyuznya 84/32, 117997 Moscow, Russia\\
  $^2$Max-Planck Institute f\"ur Radioastronomie,\\
  Auf dem H\"ugel 69, 53121 Bonn, Germany\\
  $^3$Special Astrophysical Observatory of RAS,\\
  Nizhnij Arkhyz, 369167 Russia\\
  ykovalev@avunda.asc.rssi.ru}
\\[2mm]
\end{center}
We present results of observations of 1--22 GHz instantaneous continuum
spectra of about 3000 active galactic nuclei performed in 1997--2006 with
the 600 meter ring transit radio telescope \mbox{RATAN--600}. An analysis of
types and structure of the measured instantaneous spectra has lead us to
a conclusion that almost all spectra could be modeled as a sum of two
main spectral components: LF-component (decreasing with frequency) and
HF-component (with a maximum in cm-mm band). In the framework of a model
with longitudinal magnetic field, the HF-component is explained by
synchrotron radiation of a continuous compact relativistic jet emerging
from the nucleus, the LF-component --- by radiation of optically thin
extended peripheral structures which accumulate jet particles. Long
term variability is studied in about 600 AGNs. It is dominated in the
same model by the variable emission of a compact jet (HF-component) and is
explained by variable flow of relativistic particles injected in the
jet base. We also apply another model, a standard homogeneous
blob of relativistic particles with synchrotron self-absorption, for
sources with simple parsec scale structure and peaked spectral shape. On
the basis of our combined RATAN and VLBA measurements, we estimate the
magnetic field in jet regions of these sources and compare it with
estimations provided by the model with longitudinal magnetic field.
\vspace{1cm}

\begin{center} 
{\large\bf PLANCK --- UNLOCKING THE SECRETS OF THE UNIVERSE}\\[3mm]
{\sl A. L\"ahteenm\"aki$^1$, M. Tornikoski$^1$, J. Aatrokoski$^1$, E. Valtaoja$^2$}\\[2mm]
{$^1$Mets\"ahovi Radio Observatory, Helsinki University of Technology\\
  Mets\"ahovintie 114, 02540 Kylm\"al\"a, Finland\\ 
  alien@kurp.tkk.fi}\\[2mm] 
{$^2$Tuorla Observatory, University of Turku, Finland}\\[2mm]
\end{center} 
The Planck satellite is a European Space Agency ESA's mission capable
of mapping the whole sky at several radio wavelengths. The ultimate
purpose of the satellite is to measure, with a high resolution, the
cosmic microwave background (CMB) anisotropy pattern, and thus define the
geometry and content of our Universe.  At the same time all foreground
radio sources in the sky, including extragalactic radio sources, will
be measured, too. The by-products of the CMB map cleaning process,
the foreground source maps, will become useful scientific results in
themselves. Hence the task is two-fold. First, to provide the cosmologists
with tools for cleaning the CMB maps, and second, to extract scientific
information out of the high radio frequency all-sky foreground source
catalogs.

One of the most important goals of our Planck project is the acquisition
of complete sky surveys at several high radio frequencies ---an
unprecedented event that should solve at least some of the open
questions regarding active galactic nuclei (AGNs). Even though we
do have a general idea of the basic structure and nature of AGNs, the
detailed structure and precise physical processes at work are not yet well
understood. AGNs emit at all electromagnetic frequencies from the radio
to the gamma-ray region, and all these frequencies are connected, each
frequency adding to the complete picture. The future of the AGN research
is in multi-frequency studies performed with sophisticated ground-based
and space-borne instruments or instrument networks, and Planck will be
a significant contributor to this work.

The Mets\"ahovi and Tuorla Planck team has developed a special
software called the Quick Detection System (QDS), that will be used for
detecting strong, possibly flaring, radio sources in the time-ordered
data stream of the Planck satellite within one week from the time of
observation. This is essential for follow-up observations since the actual
data product of the satellite will not be available until after two years
after the mission has started, and even the Early Release Compact Source
Catalog (ERCSC) will be available only approximately nine months after the
first full sky observation cycle has been completed. The QDS will give
us a unique opportunity to get our hands on the Planck foreground data
months before anybody else, to trigger virtually simultaneous follow-up
observations of interesting events, and also help monitor the quality of
the satellite data at an early stage. QDS will be operated in the Planck
Low Frequency Instrument (LFI) Data Processing Centre (DPC) in Trieste,
Italy, by our team. The launch date of the Planck satellite is currently
set for early 2008, and the operation of the QDS will start as soon as
the test period of the satellite has been completed.

In this paper we describe the Planck mission; the instruments and the
science it has been designed to study. A special emphasis will be made
on the Finnish participation in the project. This includes, for example,
the 70 GHz receivers that were designed and build in Finland, and
many aspects of the science we are currently involved in.
\vspace{1cm}

\begin{center}
{\large\bf POLARIZATION OBSERVATIONS OF mCVs}\\[3mm]
{\sl H.J. Lehto, S. Katajainen}\\[2mm]
{ Tuorla Observatory and Department of Physics\\
  FIN-21400 University of Turku\\
  hlehto@utu.fi}\\[2mm]
\end{center}
We have observed magnetic variable stars in polarized light in UBVRI with
the NOT.  We will discuss some of our recent results.
\vspace{1cm}
\newpage
\begin{center}
{\large\bf RADIO SPECTRA PROPERTIES OF A COMPLETE SAMPLE OF SOURCES NEAR THE NORTH CELESTIAL POLE}\\[3mm]
{\sl M.G. Mingaliev$^1$, M.G. Larionov$^2$, J.V. Sotnikova$^1$,
N.N. Bursov$^1$, N.S. Kardashev$^2$}\\[2mm]
{$^1$\it {Special Astrophysical Observatory of RAS, Nizhnij Arkhyz,
369167,  Russia; }} \\
{$^2$\it {Astro Space Center, Lebedev Physical Institute of RAS,
Moscow, Russia }}\\   marat@sao.ru\\[2mm]
\end{center}
The RATAN-600 radio telescope was used to study  spectral
properties for a complete sample of 504 sources from the NVSS
catalogue near the North Celestial Pole. The main task of the
work was to determine instantaneous spectra of radio sources with
the purpose to select objects with inverted spectra near the 22
GHz frequency for subsequent investigation under the space VLBI
Project ``RadioAstron''. The high angular resolution of the project
``RadioAstron'' which is to be achieved $~10^{-6}$ arcsec
imposes strict demands to angular dimension of
sources. These must be super-compact objects with a high value of the
correlated flux. Such objects form a considerable part of objects
just with inverse and flat spectra. At present there are no
complete high-frequency catalogues of such objects up to low flux
density levels (0.2 Jy at 22 GHz). The only available data on the
North Celestial Pole are the VLA survey (NVSS) at 1.4 GHz. It is
important to obtain spectral characteristics up to the highest
frequency of 22 GHz planned in the work of the space
interferometer. The following criteria were used in selection of
sources from the catalogue NVSS:

\begin{enumerate}
\item{$  00^{\rm h}\le RA2000 \le 24^{\rm h} $}
\item{$ +75^o\le DEC2000 \le+88^o $}
\item{Flux density: $ S_\nu[1.4{\rm GHz}]\ge 200~{\rm mJy}$ from the NVSS catalog}
\end{enumerate}

The total number of sources is 504. After data reduction
we obtained flux densities of sources and their
spectral characteristics. The sources spectral types were
determined: 65\% -- normal, 24\% -- steep, 7.3\% -- flat, 2.3\% --
inverted, and 1.4\% -- spectra with a maximum at centimeter
wavelengths (GPS).  Eleven sources with inverted spectra were
detected. The statistics of the sources spectra from our sample
contrasts with spectral characteristics of the sample of
objects with the same initial parameters but carried out at the
frequency 20 GHz by Sadler et al. (2006). We
obtained that  there is a 25\% deficit of sources with the inverted
spectra in our sample. This can be explained by the spectral properties of the
``subliminal'' sources, which did not fall into the initial sample
at the frequency of 1.4 GHz.
\\[3mm]\indent
{\bf References\\[1mm]}
Sadler E.M., Ricci R., Ekers R. D.,
Ekers J. A., Hancock P.J., Jackson C. A.,
Kesteven M.J., Murphy T., Phillips Ch.,
Reinfrank R. F., Staveley-Smith L.,
Subrahmanyan R., Walker M. A., Wilson W.E., de Zotti, G.:
{\it MNRAS},  {\bf 371},  898.
\vspace{1cm}

\newpage

\begin{center} 
{\large\bf ON THE THEORY OF RESONANT TRANSITIVE RADIATION OF DECIMETRIC RADIATION OF FLARES}\\[3mm]
{\sl E.V. Modin,  L.V. Yasnov}\\[2mm]
{Saint-Petersburg State University \\ 
  198904 Saint-Petersburg, St. Peterhof, Ul'anovskaya street 1,  Russia\\ 
  Modin.Egor@gmail.com}\\[2mm] 
\end{center} 
In this work a mechanism of resonant transitive radiation (RTR) with
reference to its possible application for interpretation of decimetric
radio emission of solar flares is analyzed. Such radiation depends
on a number of parameters of the radiating media. In particular, on
the parameter of spectrum of small-scale inhomogeneity of electronic
density, $\nu$. Platonov \& Fleishman (2002) derived the formulas
for factors RTR in dependence on the frequency of radiation for $\nu=2$. On
the whole these formulas describe the behavior of RTR precisely, however in
narrow frequency intervals they can give either negative or infinite
values. In this work, using the approaches similar to those developed
by Platonov \& Fleishman (2002),
factors of RTR for an arbitrary parameter   have been obtained.
These factors, in particular, did not give negative and infinite
values. On their basis the RTR factors integrated on frequency have been
obtained. These factors were used for the analysis of decimetric
radiation of the flare on December 24, 1991. It has been shown, that
the RTR of this flare could originate in  plasma with small-scale
inhomogeneities with $\displaystyle\frac{\left\langle\Delta
N^2\right\rangle}{N^2}=2.5\cdot10^{-5}$.
\\[3mm] \indent
{\bf References\\[1mm]}
Platonov K.Yu., Fleishman G.D.: {\it UFN}, 2002, {\bf 172}, 3, 241.
\vspace{1cm}

\begin{center} 
{\large\bf SPECTRAL ENERGY DISTRIBUTIONS AND 37 GHz MONITORING OF BL LACERTAE OBJECTS}\\[3mm]
{\sl E. Nieppola$^1$, M. Tornikoski$^1$, A. L\"ahteenm\"aki$^1$, E. Valtaoja$^2$}\\[2mm]
{  $^1$Mets\"ahovi Radio Observatory \\
  Mets\"ahovintie 114, 02540 Kylm\"al\"a, Finland\\ 
  $^2$Tuorla Observatory \\
  V\"ais\"al\"antie 20, 21500 Piikki\"o, Finland \\
  eni@kurp.hut.fi}\\[2mm]
\end{center} 
BL Lacertae objects (BL Lacs) are a group of active galactic nuclei
(AGN) characterized by strong and rapid variability, strong optical
polarization and a lack of prominent emission lines in their
spectra. We have determined spectral energy distributions (SED) for over
300 of these objects using archival multi-frequency data and fitted a
parabolic function to the synchrotron component of the SED (Nieppola
et al. 2006). The peak frequencies of the synchrotron components
range between log\,$\nu_{peak}$=\,12.67--21.46. We divided the sample
into low-energy (LBLs), intermediate energy (IBLs) and high-energy
(HBLs) BL Lacs according to their log\,$\nu_{peak}$. The correlation
between log\,$\nu_{peak}$ and the luminosity at $\nu_{peak}$ was not
significant, in contradiction with the ``blazar sequence'' scenario
(Fossati et al. 1998). We also report a summary of the first 3.5 years
of observations with the extensive BL Lac sample at 37 GHz. The BL Lac
source list contains 398 sources, all of which were observed at least
once. Roughly 34\% of the sample was detected at $S/N\,>\,4$. Most of
the detected sources were LBLs, being intrinsically more luminous at
radio wavelengths than HBLs.

{\it Acknowledgements.} The authors made use of the database CATS
(Verkhodanov et al. 1997) of the Special Astrophysical Observatory.
\\[3mm]\indent
{\bf References\\[1mm]}
Fossati, G., Maraschi, L., Celotti,
A., Comastri, A. and Ghisellini, G.: {\it MNRAS}, 1998, {\bf 299}, 433.\\
Nieppola, E., Tornikoski, M. and Valtaoja E.: {\it A\&A}, 2006, {\bf 445},
441. \\
Verkhodanov, O.V., Trushkin, S.A., Andernach, H. and Chernenkov,
V.N.: {\it ASPC}, 1997, {\bf 125}, 322.
\vspace{1cm}

\begin{center}
{\large\bf DEEP SKY SURVEYS WITH RATAN-600 }\\[3mm]
{\sl Yu.N. Parijskij}\\[2mm]
{  Special Astrophisical Observatory of RAS \\
   Nizhnij Arkhyz, 369167, Russia\\ par@sao.ru}\\[2mm]
\end{center}
Blind Sky surveys with RATAN-600 were suggested  by the general PROJECT of
AVP (1968). A Flat mirror was included into the main CIRCLE structure
to carry out quick all-sky survey, as successfully was made
with the Kraus telescope of Ohio University. In 1960th
the Sky seemed to be filled by first generation young
objects with inverted and SSA spectra, which were missing in the meters
wavelengths catalogs.  The Sternberg Astronomical Institute of the Moscow
State University group have surveyed the sky zone DEC:0-14deg
with sub-Jy sensitivity at the 2-8 cm wavelength and the first big (8500) list
of objects detected at 4 GHz objects was published just before the famous 87GB catalog
appeared (see the Zelenchuk survey catalog in CATS data base).
The CMB anisotropy studies were early on included in the high priority
scientific targets.
The first deep blind sky survey was done at 4 cm wavelength in the winter
1975-1976 with sub-mK sensitivity, but only results interesting for
CMB people were published, and they reject all available in 70-th
variants of the theories of galaxies formation. The second  epoch of
deep blind surveys started after installation of the world best
cryo-receiver at 7.6 cm (with $\sim$2mK/s$^{1/2}$ sensitivity).  The first
17 Feb. 1980 24h- drift scan demonstrated that about 200 details
may be found on this record and may be classified as radio sources, and
we integrated point sources (PS) and CMB anisotropy tasks in the experiment.
Several regions were selected for deep surveys, including the celestial Pole,
the Declination of SS433 strip, and the Declination of 3C84
(near the RATAN zenith) strip. Weakness of the CMB
anisotropy requires a great averaging (hundreds daily scans) and
we observed some regions during many years. At all frequencies lower 10
GHz we see a confusion limit, and we proposed ways to suppress this noise
using specific shape of the RATAN-600 beam. It helps us to reach the  few
mJy level at cm wavelengths and much below by P(D)  analysis. The multi-frequency
mode of observations, important for SCREENS (foregrounds) cleaning in
the CMB experiments, turned out to be useful for a spectral classification of
the PS appearing on the scans.
Now it is clear for all CMB groups, that the depth of
the CLEANING from PS is the real limit of CMB dedicated experiments,
including the PLANCK mission. The problem with PS objects at CMB frequencies
connects with an absolutely unknown Source Population (between IRAS and
NVSS, or GB). At RATAN-600 we try to use SELF CLEANING mode, using much
higher resolution than required by CMB physics (sub-degree scales). A huge
amount of data collected during the CMB experiments should be used by
PS people.  I shall present the positive and negative experience, connected
with the international BIG TRIO program, and problems with detection of a new
population at cm wavelengths. The present state of RATAN-600  ``Cold''
and RATAN-600 Zenith Field (RZF) blind surveys will be mentioned, as well
as the present-day situation with High Frequencies Sky Surveys.
This presentation
describes results of several groups in SAO and in SPb-branch of
SAO, and now is partially supported by the RFBR grant 05-02-17521, OF RAS,
SPb Center of RAS.
\vspace{1cm}

\begin{center}
{\large\bf MULTI-FREQUENCY OBSERVATIONS OF THE POLAR RADIO STRUCTURES}\\[3mm]
{\sl A. Riehokainen$^1$,  A.G. Tlatov$^2$}\\[2mm]
{$^1$Tuorla Observatory 21500, Piikkio, Finland;\\
 $^2$Pulkovo Astronomical Observatory, St. Petersburg, Russia\\
 alerie@utu.fi}\\[2mm]
\end{center}
In this work we present a comparison of the enhanced temperature regions
(ETRs) in the radio emission of the Sun with other manifestations of
solar polar structures over some days in 2003-2005. The radio observations
at 37 GHz were made  with the Metsahovi Radio Telescope (Finland). We
compared our radio data with different SOHO/EIT and SOHO/MDI images
for the same periods. We also superposed the intensity contours of the
full-radio maps obtained in Metsahovi on the Meudon Spectroheliograph
CaII(k3) and H(alpha) images. We tried to find difference between ETRs
inside and outside of coronal holes. We find that the ETRs are clearly
connected to brightness structures seen in the CaII(k3)/H$\alpha$ and
magnetic field sources seen in SOHO/MDI. Thus we can conclude that ETRs
have chromospheric origin.
\vspace{1cm}

\begin{center}
{\large\bf  NEW OBJECTS FROM THE ``COLD'' SURVEY}\\[3mm]
{\sl N.S. Soboleva, A.V. Temirova, N.N. Bursov, Yu.K. Zverev}\\[2mm]
{Special astrophysical observatory of RAS, Saint-Petersburg, Russia\\
 adelina.temirova@mail.ru }\\[2mm]
\end{center}
Results of deep surveys of a $\pm 10'$ strip of the sky centered on the
declination of SS433 carried out on the Northern sector of the RATAN-600
telescope at 2.7 and 7.6 cm wavelength in 1987-2000 are discussed.
About 600 objects at the 7.6 cm wavelength were identified with NVSS sources.
Eighteen sources detected at 2.7 cm
were not detected at 7.6 cm but could be identified with NVSS objects. It
cannot be ruled out that some of these are sources with inverted spectra. At
both wavelengths there is a fairly large number of Gaussian profiles which are
not identified with NVSS objects (106 at 2.7 cm and 43 at 7.6 cm); it is quite
possible that not all of these cases are false.

The survey objects are cross-identified with sources in the NVSS catalog
and the corresponding two-frequency spectral indices determined. We find
a decrease in the mean spectral index in the transition from objects
with flux densities $S_{21}\ge 30$ mJy to those with $15<S_{21}<30 $mJy.
The constructed $\log{N}-\log{S}$ relation at 2.7 cm has
a slope of 3/2 at flux densities $\ge300$~mJy and flattens at weaker
flux densities. The 1.4 GHz (NVSS), 3.94 GHz (RATAN-600), and 11.11 GHz
(RATAN-600) data are used to estimate the number of objects per square
degree at a wavelength of 1 cm.
\vspace{1cm}

\begin{center}
{\large\bf  REVERSAL BACKGROUND MAGNETIC FIELD IN THE SOLAR POLARIZED RADIO EMISSION AT 17 GHz }\\[3mm]
{\sl A.G. Tlatov$^1$, A. Riehokainen$^2$}\\[2mm]
{$^1$Kislovodsk  Solar Station of the Pulkovo Astronomical Observatory, Russia \\
 $^2$Tuorla Observatory, Turku University, Finland \\
  solar@narzan.com}\\[2mm]
\end{center}
Polarization of radio emission on the solar disk was studied
with the Nobeyama radio heliograph observations during
1992-2006. The latitude-time diagrams of polarization circular
radio emission were constructed. To decrease the noises we
used several solar images for a day. We found  drifts
of radio emission polarization in the high-latitudes activity and in the latitude
band of sunspots. Process of the magnetic field reversal of
the large-scale magnetic field in polarization of radio emission
of the Sun was found during 22-23 cycles. An analysis of
polarization for structures of various brightness temperatures
has been carried out.
\vspace{1cm}

\begin{center}
{\large\bf OSCILLATION OF THE POLARIZED RADIO EMISSION  ``THE SUN AS A STAR'' }\\[3mm]
{\sl A.G. Tlatov$^1$,  A. Riehokainen$^2$}\\[2mm]
{$^1$Kislovodsk  Solar Station of the Pulkovo Astronomical Observatory, Russia \\
 $^2$Tuorla Observatory, Turku University, Finland \\
  solar@narzan.com}\\[2mm]
\end{center}
We investigated variations of the radio emission of the whole Sun
at 1.76 cm wavelength obtained and archived at the Nobeyama radio heliograph
in 1992-2006. For this purpose the daily data of the intensity
and also right/left circular polarization of the radio emission
with one-second average were processed. It was found that 3
minutes oscillations are present at the different phases of solar
activity, including the minimum of activity. Especially
conspicuous these oscillations present in a difference between the
right and the left circular polarization. Intensity of the
oscillations changes with a level of the solar activity. Spectral
analysis of the presence of 3-minute oscillations in polarization
of the solar radio emission shows that there exist a modulation with
the periods of 27 and 157 days. During the minimum activity the
main periods of the 3 minutes oscillations are slightly shorter
than during the maximum activity.
\vspace{1cm}

\begin{center} 
{\large\bf RADIO SPECTRA OF GPS GALAXIES}\\[3mm]
{\sl I. Torniainen$^1$, M. Tornikoski$^1$, M. Aller$^2$, H. Aller$^2$, M. Mingaliev$^3$}\\[2mm]
{ $^1$ Mets\"ahovi Radio Observatory, Helsinki University of Technology\\
  Mets\"ahovintie 114, FIN-02540 Kylm\"al\"a, Finland\\ 
  $^2$ Department of Astronomy, University of Michigan\\
  $^3$ Special Astrophysical Observatory of RAS \\
  ilo@kurp.hut.fi}\\[2mm]
\end{center} 
Gigahertz-peaked spectrum (GPS) sources are active galactic nuclei
which are characterized by a convex radio continuum spectrum peaking
at the GHz-frequencies. Their nature is still unclear, but currently
the strongest scenario suggests that at least some of them are newborn
radio sources in which the activity has been triggered on only 100 --
1000 years ago.  There are both quasar and galaxy type GPS sources, which
have a similar shape of spectrum but the nature and the physics of
sources are thought to be different.  Our earlier study (Torniainen
et al. 2005) showed that a considerable proportion of quasar-type GPS
sources are more likely misidentified flat-spectrum quasars -- not GPS
sources at all.

We have collected 96 GPS galaxies from the literature, observed them and
collected all possible radio data for them to study how pure the galaxy
type GPS samples are.  Our sample includes both frequently monitored
sources and sources with only a few detections.  The spectra of the
sample show that less than a third of our sample were definitely or highly
probably GPS sources whereas less than a third did not have enough data
for any solid classification. Five sources had a convex spectrum but
high variability and the rest had steep or flat spectrum.  These results
show that the GPS galaxy samples have more genuine GPS sources than the
quasar samples but yet a remarkable share of them cannot be classified
as GPS sources.

Difference between the quasar and galaxy samples can partly be
explained by selection effects: the quasar sample was selected from the
Mets\"ahovi monitoring sample which has been monitored over 25 years
whereas the galaxy sample was gathered from the GPS literature and
included both weak or rarely observed sources and more frequently
monitored sources.

{\it Acknowledgements.} The authors made use of the database CATS
(Verkhodanov et al. 1997) of the Special Astrophysical Observatory.
\\[3mm]\indent
{\bf References\\[1mm]}
Torniainen I., Tornikoski
M., Ter\"asranta H., Aller M.F., Aller H.D.: {\it A\&A}, 2005,
{\bf 435}, 839. \\
Verkhodanov O.V., Trushkin S.A., Andernach H.,
Chernenkov V.N.: {\it ASP Conference Series}, 1997, {\bf 125}, 322.
\vspace{1cm}

\newpage
\begin{center} 
{\large\bf METS\"AHOVI AGN PROJECTS CONTRIBUTING TO THE PLANCK FOREGROUND SCIENCE}\\[3mm]
{\sl M. Tornikoski$^1$, A. L{\"a}hteenm{\"a}ki$^1$, T. Hovatta$^1$,
E. Nieppola$^1$, I. Torniainen$^1$, E. Valtaoja$^2$}\\[2mm]
{$^1$Helsinki University of Technology, Mets{\"a}hovi Radio Observatory \\
  Mets{\"a}hovintie 114, 02540--Kylm{\"a}l{\"a}, Finland\\ 
 $^2$Tuorla Observatory, University of Turku, Finland \\
 merja.tornikoski@tkk.fi}\\[2mm]
\end{center} 
During recent years we have had a special focus in our Mets{\"a}hovi
observing projects. We have put an emphasis on the understanding 
of  AGNs that could contribute to the extragalactic foreground
that will be detectable by the Planck satellite. 
 
First of all, we have observed completely new source samples. 
Many AGN samples have been excluded from high-frequency radio 
observations earlier simply because they were assumed to be 
too faint or ``uninteresting''. One of our largest new source 
samples was the complete BL Lacertae Object (BLO) sample. 
 
In addition to the few-epoch observations of large source 
samples we have been interested in the long-term variability 
behaviour of a densely monitored set of sources. We have 
analysed these data in order to improve our understanding 
of the variability behaviour of these sources: how often 
do flares typically occur in a certain source, and how likely 
is e.g. the Planck satellite to detect a source in  
a flaring state at a random observing epoch? 
 
We are also working our way towards predicting, or 
at least making ``educated guesses'' about, the activity 
behaviour of radio-bright AGNs. 
 
In this presentation we will discuss our source samples 
and show some recent results. 
\vspace{1cm}

\begin{center}
{\large\bf  NEW WMAP CATALOG SOURCES OR HOW MANY BRIGHT SOURCES ARE ON THE SKY}\\[3mm]
{\sl S.A. Trushkin}\\[2mm]
{  Special Astrophysical Observatory of RAS,  Nizhnij Arkhyz,  Russia\\
  satr@@sao.ru}\\[2mm]
\end{center}
We continued studies of the WMAP-sources after publishing
results of three years (Hinshaw et al.; Jarosik et al.; Page et al.;
Spergel  et al. 2006).
Trushkin (2003) presented compiled radio spectra of 205 extragalactic sources
from the catalog, compiled from the WMAP survey data
at 23-94 GHz in the first year of its operation.

We have shown that 205 WMAP-sources are reliably identified
with radio sources from known catalogs (FIRST, NVSS and so on).
$\sim$50\% of the sources have flat or inverted spectra,
$\sim$15\% -- spectra with the peaks at 5-20 GHz (GPS-sources),
$\sim$10\% -- power-law spectra and
$\sim$10\% -- show the composite spectra (as 3C84).

We discussed recent results of the radio observations of the AGNs from
the WMAP catalog with the RATAN-600 and RT32 telescopes.

Now the final 3-year catalog contains 323 sources,
while four identified sources from the former version were not included
in the new one.
Using again the search {\it select and match} and spectra plotting
procedures in our CATS data base (Verkhodanov et al. 1997) we have found
optical and radio identifications for the most of new 120 WMAP-sources
from the radio and optical catalogs in the CATS data base.
Now we discuss results for new 120 sources, their spectra and features.
The statistics for types of the sources did not generally changed.
We have found that 313 WMAP-sources have optical counterparts:
220 -- quasars, 30 -- galaxies, 32 -- AGNs,
30 -- BL Lac objects and one -- the planetary nebula IC418.
We have observed some of the new WMAP-sources  with RT32 (IAA)
at 2.3 and 8.5 GHz.
Comparison fluxes from the first version catalog and second one
allows us to estimate of the variability of the sample.
As was expected, the index of the variability at 63 GHz is higher than
at 23 GHz. There are the sources with 100\% changes of the fluxes on
the effective time scale about one year, (3yr-1yr)/2 in the WMAP3-catalog.

Using an analogous method of the source selection from the CATS data base
we have found that a probable number of the sources
brighter 400 mJy at 23 GHz is equal to 1300-1500 on the sky.
Thus there is a strong effect of confusion dramatically
decreasing the number (323) of detected sources in the WMAP survey.

We hope that such studies will help in the future PLANCK CMB-experiment
data-precessing.

{\it Acknowledgements.}
We are thankful to  RFBR for support, the grant N05-02-17556.
\\[3mm]\indent
{\bf References\\[1mm]}
Hinshaw G. et al.: astro-ph/0603451. \\
Jarosik N. et al.: astro-ph/0603452.  \\
Page L. et al.:    astro-ph/0603450. \\
Spergel D.N. et al.: astro-ph/0603449. \\
Trushkin S.A.: {\it Bulletin of SAO RAS}, 2003, {\bf 55}, 90.\\
Verkhodanov O.V., Trushkin S.A., Andernach H., Chernenkov V.N.:
{\it ASP Conference Series}, 1997, {\bf 125}, 322.
\vspace{1cm}

\begin{center}
{\large\bf RECENT DATA OF THE MULTI-FREQUENCY MONITORING OF MICROQUASARS }\\[3mm]
{\sl S.A. Trushkin }\\[2mm]
{Special Astrophysical Observatory of RAS, Nizhnij Arkhyz, Russia\\
   satr@sao.ru}\\[2mm]
\end{center}

We discuss results of recent radio observations of the microquasars
SS433, GRS1915+105 and Cyg X-3 with the RATAN-600 and RT32 (IAA)
radio telescopes.

We have carried out long monitoring programs
of daily observations sets for microquasars with the RATAN radio
telescope at frequencies of 1, 2.3, 4.8, 7.7, 11, 21.7 and 30 GHz.
Flaring events were detected when the fluxes increased  by a factor
of 2-200. The flaring synchrotron emission indicates the jets formation,
coupling with accretion disk activity in the Galactic microquasars and
in the active galactic nuclei (AGN).
The multi-frequency light curves are compared with the XTE ASM data at 2-12 keV
to study correlations during the flares. Indeed in many cases such
correlations were detected.

Radio flaring events of SS433, Cyg X-3, LSI+61d303 were often
optically thin at $\nu>$5 GHz, and follow to general predictions
of the relativistic outflows of mass or fast electrons from binaries.
On the other hand, we have often measured the inverted (optically thick)
spectra of flaring events in the active states of GRS1915+105, Cyg X-3
and V4641~Sgr with the spectral indices $\alpha>$+1 at $\nu>$1 GHz.

We have detected a clear X-ray/radio association in light curves
of GRS 1915+105 during October-November 2005, when it was very active
(0.5-3 Crabs at 2-12 keV).

After 18 days of the quenched state ($\sim$10 mJy) Cyg X-3 exhibited the 1Jy-
radio flare on 1 Feb 2006. Such a remarkable property -- before a flare
the radio emission fall down in deep (local) minimum of fluxes -- is probably
a general feature of the radio/X-ray binaries.

The flare of 1 Feb was also detected with the Nobeyama 45m and
NMA telescopes (Tsuboi et al. 2006), and for the first time a flat radio
spectrum of the flaring event from Cyg X-3 was directly measured in
the quasi-simultaneous observations from 2 to 110 GHz.  Then two following
flaring  events (5 and 17 Jy) were detected later during $\sim$100 days.
Their durations were ~50 and ~30 days respectively.
The very fast rising flare, from 1 to 2 Jy during  3 hours,  was  detected
with the RT32  telescope  (Trushkin et~al. 2006) on 05  June.
At last on 25 July we have detected a very powerful flare (15 Jy) from
Cyg X-3 again. All these flares happened during a long period (Feb 1 -- Aug 1)
when X-ray emission was relatively high ($\ge$0.3 crabs), variable and hard.

We studied evolution of the powerful flares from the optically thick state
to the optically thin one at the lower frequencies. We have to draw an
unexpected conclusion: during the stage of initial rising (ejection stage)
the density of thermal electrons is also rising resulting in the higher
optical depths at frequencies lower than 1 GHz just near maximum of the flare.

{\it Acknowledgements.}
This research was supported by the Russian Foundation of Basic Research
grant N 05-02-17556 and  by the Program of the Presidium of  Russian
Academy of Science.
\\[3mm]\indent
{\bf References\\[1mm]}
Tsuboi M. et al. :    {\it ATel}, 2006, {\bf \#729}.\\
Trushkin S.A. et al.: {\it ATel}, 2006, {\bf \#828}.
\vspace{1cm}

\begin{center}
{\large\bf THE PROBLEM OF HIGH-ENERGY EMISSION FROM AGN}\\[3mm]
{\sl Esko Valtaoja}\\[2mm]
{  Tuorla Observatory, University of Turku, FI-21500 Piikkio, Finland\\
esko.valtaoja@utu.fi}\\[2mm]
\end{center}
The basic framework for radio-bright AGNs, which are also the only types
of extragalactic sources known to emit a significant amount of high-energy
radiation, is a relativistic jet with shocks embedded in it. Presumably,
the most significant intrinsic properties of the source are then the
absolute luminosity of the jet/shocks, the flow speed (the Lorentz
factor) and the angle which the jet makes to the line of sight. The more
detailed nature of the flow (accelerating/decelerating flow, turbulence,
particle acceleration/reacceleration, magnetic field configuration,
jet opening angle and curvature, duty cycle of the shock activity, etc.)
as well as the jet surroundings (in particular, density of the ambient
photon field) must also play a role.\\
The spectral energy distributions
of radio-bright AGN, often called blazars, can be approximated by two
parabolas. The first one is caused by synchrotron radiation from the jet
and from the shocks, the second one by an inverse Compton radiation from
the relativistic electrons in the jet, upscattering ambient photons
into X-to-TeV energies.\\
Both theoretically and observationally, our
understanding of the blazar emission remains rather poor. I discuss some
new attempts to model the spectral energy distributions of blazars,
focusing on correlations between various observed and intrinsic
properties, and on the problems of the proposed theoretical models for
the high-energy emission.
\vspace{1cm}

\begin{center}
{\large\bf OPEN WEB-RESOURCES OF SAO RAS FOR EXTRAGALACTIC RESEARCH}\\[3mm]
{\sl O.V. Verkhodanov $^1$, S.A. Trushkin $^1$,  A.I. Kopylov$^1$, V.N. Chernenkov$^1$,
H. Andernach $^2$, N.V. Verkhodanova $^1$, V.K. Kononov $^1$
}\\[2mm]
{
$^1$Special Astrophysical Observatory of RAS, Nizhnij Arkhyz,  Russia\\
$^2$  Departamento de Astronom{\'{\i}}a, Universidad de Guanajuato, Mexico\\
 vo@sao.ru}\\[2mm]
\end{center}
The sky survey at various wavelength bands is one of the most important
branches of the modern observational astrophysics.
To the present moment, several data bases unify resulting data of such
surveys. We consider here  Web-resources designed for radio astronomical
and extragalactic study and operating in the Special Astrophysical
Observatory. We describe  structure, operation, standard tasks and
the current status of the largest data bases in  respective astrophysics
fields. They are the data base of radio astronomical and astrophysical
catalogs CATS ({\tt http://cats.sao.ru}) (Verkhodanov et al. 1997, 2005)
 and the SED (Verkhodanov et al. 2000) which is an analysis system of
spectral energy distribution ({\tt http://sed.sao.ru}). These servers
were unified in a one cluster and are used both for separate source and
statistical studies in astrophysics and cosmology. A new site
({\tt cmb.sao.ru}) is prepared for cosmology. It was worked out with the same
approach as the above mentioned. It bases on the GLESP (Doroshkevich et al. 2005)
package and allows a user to transform spherical harmonics
to full sky maps. It contains the WMAP CMB and foreground maps and
corresponding files with $a_{\ell m}$ coefficients of spherical harmonics.
We plan to unify all three servers into one cluster for acceleration
of investigation in the field of cosmology and astrophysics.

{\it Acknowledgements.}
This work is supported particularly by the Russian Foundation of Basic
Research (grant No 05-07-90139).
\\[3mm]\indent
{\bf References\\[1mm]}
Doroshkevich A.G., Naselsky P.D., Verkhodanov O.V., Novikov D.I.,
 Turchaninov V.I., Novikov I.D., Christensen P.R., Chiang L.-Y.:
 {\it Internat. J. Mod. Phys. D}, 2005, {\bf 14}, 275 (astro-ph/0305537) \\
Verkhodanov O.V., Trushkin S.A., Andernach H., Chernenkov V.N.:
    In ``Astronomical Data Analysis Software and Systems VI'',
    eds. G.Hunt \& H.E. Payne, 1997, {\it ASP Conf. Ser.}, {\bf 125}, 322
    (astro-ph/9610262) \\
Verkhodanov O.V., Kopylov A.I., Zhelenkova O.P., Verkhodanova N.V.,
 Chernenkov V.N., Parijskij Yu.N., Soboleva N.S., Temirova A.V.:
 The software system ``Evolution of radio galaxies''.
 {\it Atsron. Astrophys. Trans.}, 2000, {\bf 19}, 662, (astro-ph/9912359) \\
Verkhodanov O.V., Trushkin S.A., Andernach H., Chernenkov V.N.:
    {\it Bulletin of SAO of RAS}, 2005, {\bf 58}, 118
\vspace{1cm}

\begin{center}
{\large\bf PHASE ANALYSIS IN STUDY OF COSMIC MICROWAVE BACKGROUND}\\[3mm]
{\sl O.V.Verkhodanov$^1$, P.D.Naselsky$^2$, L.-Y.Chiang$^2$, A.G.Doroshkevich$^3$,
I.D.Novikov$^{3,2}$
}\\[2mm]
{
$^1$Special astrophysical observatory  of RAS, Nizhnij Arkhyz,  Russia\\
$^2$Niels Bohr Institute, Copenhagen, Denmark  \\
$^3$AstroSpace Center, Moscow, Russia \\
 vo@sao.ru}\\[2mm]
\end{center}
Phase analysis based on the complex values of spherical harmonics
of the CMB fluctuation expansion is an extremely important method of
observational cosmology. We consider several aspects of phase analysis
in the CMB study. They concern the problems of signal restoration (Naselsky
et al. 2005), search for non-Gaussianity (Chiang et al. 2003) and study of
the foreground contamination of this separated CMB signal (Naseslky et al.
2003, 2004, 2005). Using the GLESP (Gauss-LEgendre Sky Pixelization)
package (Doroshkevich et al. 2005) for CMB analysis we produce phase data
for spherical harmonics in the form of
$a_{\ell m} = |\delta_{\ell m}|\exp(i\Psi_{\ell m})$,
where $a_{\ell m}$-s are  coefficients of spherical harmonics of
the $\ell$--multipole in the $m$--mode,
$|\delta_{\ell m}|$ is their amplitude,
$\Psi_{\ell m}$ is a phase, and $i$ is the imaginary unit.
Considering the minimum in correlation of phases of CMB and foregrounds
we can separate the signal corresponding to these properties at low $\ell$
($\ell\le100$) where the point source influence is minimal.
Another important moment is the study of the Gaussianity problem
in the CMB observational data. To check  statistical properties
of the data we produce phase diagrams. These diagrams demonstrate
the strong non-Gaussianity for all accessible maps of CMB.
The high phase correlations between CMB and foregrounds hint us about
problems of signal separation. Similar approaches are developed for
the {\rm Planck} mission.

{\it Acknowledgements.}
This work is supported particularly by the Russian Foundation of Basic
Research (grants No 05-07-90139  and  05-02-16302).
\\[3mm]\indent
{\bf References\\[1mm]}
Chiang L.-Y., Naselsky P.D., Verkhodanov O.V., Way M.J."
 {\it Astrophys. J.}, 2003, {\bf 590}, L65 (astro-ph/0303643) \\
Naselsky P.D., Doroshkevich A.G. , Verkhodanov O.V.:
 {\it Astrophys. J.}, 2003, {\bf 599}, L53  (astro-ph/0310542) \\
Naselsky P.D., Doroshkevich A.G., Verkhodanov O.V.:
    MNRAS, 2004, {\bf 349}, 695 (astro-ph/0310601) \\
Naselsky P.D., Chiang L.-Y., Novikov I.D., Verkhodanov O.V.:
 {\it Internat. J. Mod. Phys. D}, 2005, {\bf 14}, 1273 (astro-ph/0405523) \\
Doroshkevich A.G., Naselsky P.D., Verkhodanov O.V., Novikov D.I.,
 Turchaninov V.I., Novikov I.D., Christensen P.R., Chiang L.-Y.:
 {\it Internat. J. Mod. Phys. D}, 2005, {\bf 14}, 275 (astro-ph/0305537)
\\[3mm]\indent
\vspace{1cm}
\newpage
\begin{center}
{\large\bf THE COMBINED RADIO AND OPTICAL INVESTIGATIONS OF THE INTRADAY VARIABILITY OF ACTIVE GALACTIC NUCLEI}\\[3mm]
{\sl A.E. Volvach$^1$, V.S. Bychkova$^2$, N.S. Kardashev$^2$,\\
M.G. Larionov$^2$, V.V. Vlasyuk$^3$, O.I. Spiridonova $^3$}\\[2mm]
{$^1$ SRI Crimean Astrophysical Observatory, Ukraine, \\
$^2$ Astro Space Centre of the Lebedev Physical Institute, Russia\\
$^3$ Special Astrophysical Observatory of the Academy Science, Russia \\
volvach@crao.crimea.ua}\\[2mm]
\end{center}
The combined radio and optical observations of the active galactic nuclei
0133+476, 1633+382, 2145+067, 2251+158 were performed in  2004-2006.
The aim of analysis was to detect an intraday flux
variability and to search for its possible correlation in radio and
optical wavelengths. Observations were conducted with the use of the RT-22 radio
telescope (SRI CrAO) at 22 and 36 GHz and the 1-m Zeiss-1000 reflector of
SAO RAS with CCD camera.

During  observations  we found no significant fluctuations of fluxes
at both ranges.  At the level of 10\% from the average amplitude more high
activity of the object 1633+382 was detected at 36 GHz in May 2004 then
the one in May 2005. The reason of it may be a passive phase of the
object after the burst in 2002. The observable flux variation in
0133+476 at the time scale of one hour were noticed in October 2005 but
not optical range. Such flux behaviour may indicate to absence of the
identical area for radio and optical emission after matter collimation
in the black hole polar region.
\\[3mm]\indent
{\bf References\\[1mm]}
Volvach A.E.,  Larionov M.G., Aller M., Aller H.:
{\it Radio Astronomy and Radio Physics},
2005. {\bf 10}, N.4, 377.
\vspace{1cm}

\begin{center}
{\large\bf PHYSICAL AND CHEMICAL STRUCTURE OF HIGH MASS STAR FORMING REGIONS }\\[3mm]
{\sl I.I. Zinchenko}\\[2mm]
{  Institute of Applied Physics of the Russian Academy of Sciences \\
  46 Uljanov str., Nizhny Novgorod 603950,  Russia\\
  zin@appl.sci-nnov.ru}\\[2mm]
\end{center}
In recent years we surveyed several tens of high mass star forming
regions in various molecular lines and in millimeter wave continuum.
Basic physical properties of detected clumps and molecular
abundances were derived. One of the problems is a selection of the
best tracer of mass distribution. In particular, we found that
in regions of high mass star formation the CS emission correlates
well with the dust continuum emission and is therefore a good tracer
of the total mass while the N$_2$H$^+$ distribution is frequently
very different. This is opposite to their typical behavior in
low-mass cores where a freeze-out plays a crucial role in the
chemistry. The behavior of other high density tracers varies from
source to source but most of them are closer to CS. Radial density
profiles in massive cores are fitted by power laws with indices
about $-1.6$, as derived from the dust continuum emission. The
radial temperature dependence on intermediate scales is close to the
theoretically expected one for a centrally heated optically thin
cloud. The velocity dispersion either remains constant or decreases
from the core center to the edge. Several cores including those
without known embedded IR sources show signs of infall motions. They
can represent the earliest phases of massive protostars. There are
implicit arguments in favor of small-scale clumpiness in the cores.

{\it Acknowledgements.} The work was supported by the Russian Foundation
for Basic Research grant 06-02-16317 and by the Program ``Extended
objects in the Universe'' of the Russian Academy of Sciences.
\newpage

{\Large \bf Author's index}\\[3mm]
Aatrokoski       J.   \hspace{1cm} 25              \\
Abramov-Maksimov V.E. \hspace{0.3cm} 9,11            \\
Alferova         Z.A. \hspace{1cm} 20              \\
Aller            H.   \hspace{1cm} 21,32           \\
Aller            M.   \hspace{1cm} 21,32           \\
Andernach        H.   \hspace{1cm} 36              \\
Arshakian        T.G. \hspace{1cm}  9              \\
Berlin           A.B. \hspace{1cm} 14              \\
Bochkarev        N.G. \hspace{1cm} 9,12            \\
Bogod            V.M. \hspace{1cm} 11,11,18,23     \\
Borisov          N.   \hspace{1cm} 12              \\
Borovik          V.N. \hspace{1cm} 11              \\
Burenkov         A.N. \hspace{1cm} 9,12            \\
Bursov           N.N. \hspace{1cm} 14,14,18,30     \\
Bychkova         V.S. \hspace{1cm} 38              \\
Carrasco         L.   \hspace{1cm} 12              \\
Chavushyan       V.H. \hspace{1cm}  9              \\
Chernenkov       V.N. \hspace{1cm} 36              \\
Chiang           L.-Y.\hspace{1cm} 37              \\
Collin           S.   \hspace{1cm} 12              \\
Doroshenko       V.T. \hspace{1cm} 12,15           \\
Doroshkevich     A.G. \hspace{0.6cm} 37              \\
Efimov           Y.S. \hspace{1cm} 16              \\
Fabrika          S.N. \hspace{1cm} 17              \\
Garaimov         V.I. \hspace{1cm} 11              \\
Gelfreikh        G.B. \hspace{1cm}  9              \\
Golubchina       O.A. \hspace{1cm} 18              \\
Gorshkov         A.G. \hspace{1cm} 19              \\
Gosachinskij     I.V. \hspace{0.6cm} 20,20           \\
Grechnev         V.V. \hspace{1cm} 11              \\
Grigorieva       I.Y. \hspace{1cm} 11              \\
Harinov          M.   \hspace{1cm} 21              \\
Hovatta          T.   \hspace{1cm} 21,33           \\
Ili\'c           D.   \hspace{1cm} 12              \\
Kaltman          T.I. \hspace{1cm} 11,22,23        \\
Kardashev        N.S. \hspace{1cm} 38              \\
Katajainen       S.   \hspace{1cm} 26              \\
Khabibullina     M.L. \hspace{0.6cm} 14              \\
Kononov          V.K. \hspace{1cm} 36              \\
Kopylov          A.I. \hspace{1cm} 36              \\
Korzhavin        A.N. \hspace{1cm} 11,18,22,23     \\
Kotelnikov       V.   \hspace{1cm} 23              \\
Kouprianova      E.G. \hspace{1cm} 24              \\
Kovalev          Yu.A.\hspace{1cm} 25              \\
Kovalev          Y.Y. \hspace{1cm} 25              \\
L\"ahteenm\"aki  A.   \hspace{0.7cm} 25,28,33        \\
Lainela          M.   \hspace{1cm} 21              \\
Larionov         M.G. \hspace{1cm} 27              \\
Lehto            H.J. \hspace{1cm} 26              \\
Lobanov          A.P. \hspace{1cm}  9              \\
Majorova         E.K. \hspace{1cm} 14              \\
Mikhailov        A.   \hspace{1cm} 21              \\
Mingaliev        M.   \hspace{1cm} 14,19,27,32     \\
Modin            E.V. \hspace{1cm} 28              \\
Naselsky         P.D. \hspace{1cm} 37              \\
Nieppola         E.   \hspace{1cm} 28              \\
Nizhelskij       N.A. \hspace{1cm} 14,25           \\
Novikov          I.D. \hspace{1cm} 37              \\
Parijskij        Y.N. \hspace{1cm} 29              \\
Popovi\'c        L.   \hspace{1cm} 12              \\
Riehokainen      A.   \hspace{1cm} 30,31,31        \\
Semenova         T.A. \hspace{1cm} 14              \\
Sergeev          S.G. \hspace{1cm} 15              \\
Shapovalova      A.I. \hspace{0.6cm} 9,12            \\
Soboleva         N.S. \hspace{1cm} 30              \\
Sotnikova        J.V. \hspace{1cm} 27              \\
Stepanov         A.V. \hspace{1cm} 24              \\
Temirova         A.V. \hspace{1cm} 30              \\
Tlatov           A.G. \hspace{1cm} 30,31,31        \\
Tokhchukova      S.Kh.\hspace{0.5cm} 18,23           \\
Torniainen       I.   \hspace{1cm} 21,32           \\
Tornikoski       M.   \hspace{1cm} 21,25,28,32,33  \\
Trushkin         S.A. \hspace{1cm} 21,33,34,36     \\
Tsibulev         P.G. \hspace{1cm} 14              \\
Valdes           J.R. \hspace{1cm} 12              \\
Valtaoja         E.   \hspace{1cm} 21,25,28,35     \\
Venger           A.P. \hspace{1cm} 20              \\
Verkhodanova     N.V. \hspace{0.5cm} 36              \\
Verkhodanov      O.V. \hspace{0.6cm} 36,37           \\
Vlasuyk          V.V. \hspace{1cm} 12              \\
Vovk             E.Yu.\hspace{1cm} 16              \\
Yasnov           L.V. \hspace{1cm} 23,28           \\
Zaitsev          V.V. \hspace{1cm} 24              \\
Zensus           J.A. \hspace{1cm}  9              \\
Zhdanova         V.E. \hspace{1cm} 12              \\
Zhekanis         G.V. \hspace{1cm} 25              \\
Zinchenko        I.I. \hspace{1cm} 38              \\
Zverev           Yu.K.\hspace{1cm} 30              \\
\end{document}